\documentclass[compsoc]{IEEEtran}

\usepackage{cite}
\usepackage{amsmath,amssymb,amsfonts}
\usepackage{algorithmic}
\usepackage{graphicx}
\usepackage{textcomp}
\usepackage{xcolor}
\usepackage{url}
\usepackage{booktabs}
\usepackage{xspace}
\usepackage{colortbl}
\usepackage{multicol}
\usepackage{multirow}
\usepackage[export]{adjustbox}
\usepackage{makecell}
\usepackage{silence}
\usepackage[caption=false, font=footnotesize]{subfig}
\usepackage{enumitem}
\setlist[itemize]{leftmargin=0.675cm}
\usepackage{ragged2e}
\usepackage{cite}
\usepackage{pbalance}
\usepackage{listings}

\definecolor{codegray}{rgb}{0.5,0.5,0.5}
\definecolor{commentcolor}{RGB}{61,122,122}
\lstdefinestyle{mystyle}{
    language=bash,
    commentstyle=\color{commentcolor},
    keywordstyle=\color{magenta},
    numberstyle=\tiny\color{codegray},
    basicstyle=\ttfamily\footnotesize,
    breakatwhitespace=false,
    breaklines=true,
    captionpos=b,
    keepspaces=true,
    numbersep=5pt,
    showspaces=false,
    showstringspaces=false,
    showtabs=false,
    tabsize=2,
}
\newcommand{\Dalek}{\textsc{Dalek}\xspace}
\newcommand{\R}{\textsuperscript{\textregistered}\xspace}
\newcommand{\TM}{\textsuperscript{\texttrademark}\xspace}
\newcommand{\Asus}{Asus\R}
\newcommand{\Minisforum}{Minisforum\R}
\newcommand{\AMD}{AMD\R}
\newcommand{\ARM}{ARM\R}
\newcommand{\Intel}{Intel\R}
\newcommand{\Nvidia}{Nvidia\R}
\newcommand{\Microchip}{Microchip\R}
\newcommand{\Texas}{Texas Instruments\R}
\newcommand{\Noctua}{Noctua\R}

\newcommand{\Kingston}{Kingston\R}

\newcommand{\Ryzen}{Ryzen\TM}
\newcommand{\Core}{Core\TM\xspace}
\newcommand{\CoreUltra}{Core\TM~Ultra\xspace}
\newcommand{\GeForce}{GeForce\TM}
\newcommand{\Radeon}{Radeon\TM}
\newcommand{\Arc}{Arc\TM}

\begin{document}

\title{\LARGE \Dalek: An Unconventional \& Energy-aware Heterogeneous Cluster}

\author{\IEEEauthorblockN{Adrien Cassagne\IEEEauthorrefmark{2},
        Noé Amiot\IEEEauthorrefmark{2} and
        Manuel Bouyer\IEEEauthorrefmark{2}}\\
\IEEEauthorblockA{\IEEEauthorrefmark{2}{\small LIP6, Sorbonne Université, CNRS,
UMR 7606 Paris, France}}
\thanks{This study has been carried out with financial support from the French
State, managed by the French ``Agence Innovation Défense''.}%
}

\markboth{Pre-print document, July~2025}%
{A. Cassagne \MakeLowercase{\textit{et al.}}: \Dalek: A Cluster Unlike Any
  Other}

\IEEEtitleabstractindextext{%
\begin{abstract}
\Dalek is an experimental compute cluster designed to evaluate the performance
of heterogeneous, consumer-grade hardware for software design, prototyping, and
algorithm development. In contrast to traditional computing centers that rely on
costly, server-class components, \Dalek integrates CPUs and GPUs typically found
in mini-PCs, laptops, and gaming desktops, providing a cost-effective yet
versatile platform. This document details the cluster's architecture and
software stack, and presents results from synthetic benchmarks. Furthermore, it
introduces a custom energy monitoring platform capable of delivering 1000
averaged samples per second with milliwatt-level resolution. This high-precision
monitoring capability enables a wide range of energy-aware research experiments
in applied Computer Science.
\end{abstract}

}

\maketitle
\IEEEdisplaynontitleabstractindextext

\section{Introduction}

\IEEEPARstart{D}{alek} is an innovative cluster built around CPUs typically used
in mini-PCs or laptops, and GPUs commonly found in gaming PCs (or integrated
GPUs). The cluster includes a wide range of recent components that can be tested
across various algorithms. One of the main purposes of \Dalek is to provide such
hardware diversity at a moderate cost. Indeed, consumer-grade components are
significantly less expensive than server-class hardware. As a result, \Dalek is
particularly well-suited for software design and prototyping, enabling
researchers to explore and experiment with new hardware shortly after its
release.

\begin{figure}[t]
  \centering
  \includegraphics[width=0.95\linewidth]{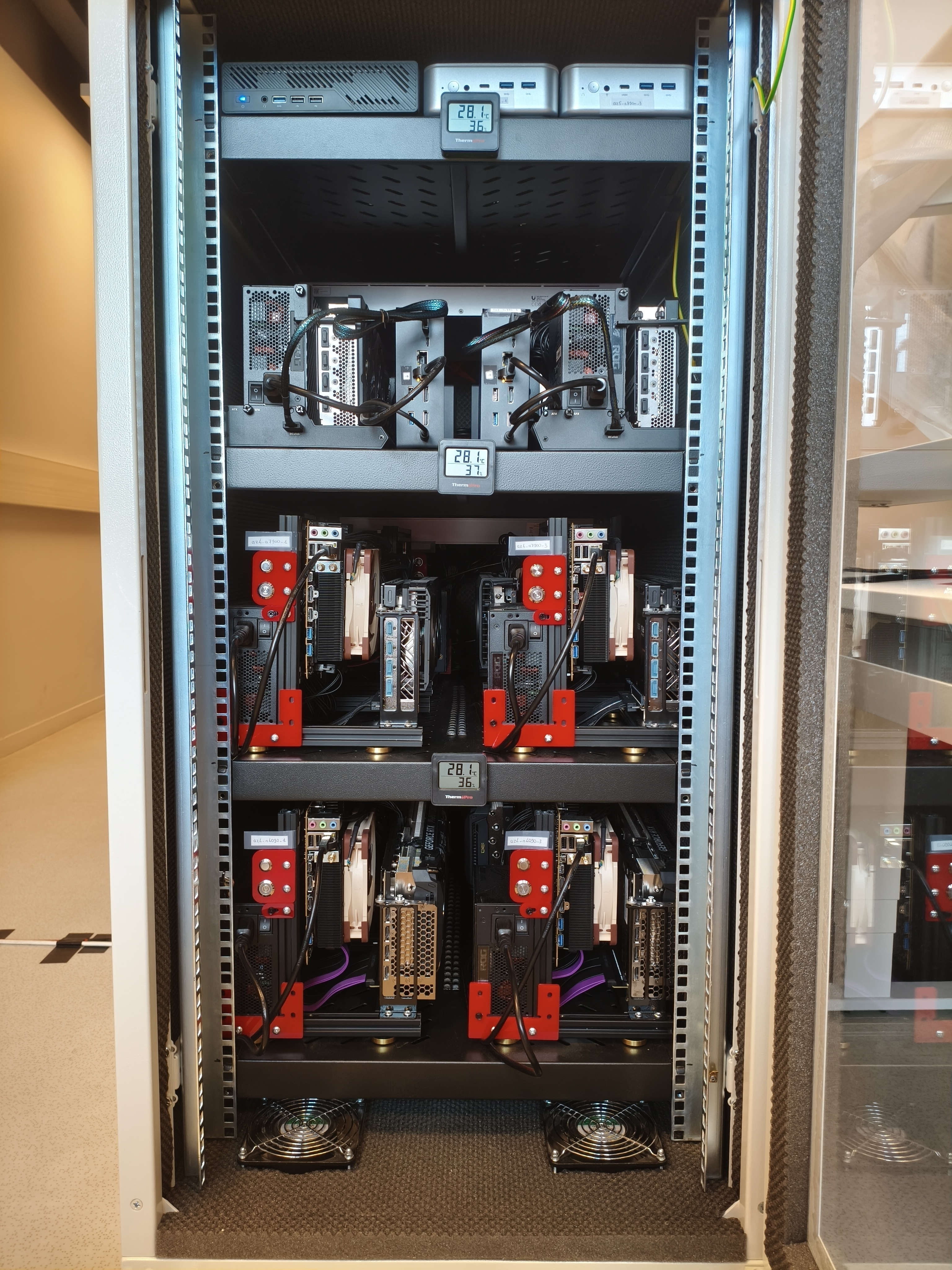}
  \caption{\Dalek cluster. Photo taken the June 21\textsuperscript{th}, 2025.
    Each level corresponds to a compute partition composed of four nodes each.
    25~U rack on wheels of 1.0$\times$0.6$\times$1.3m dimensions.}
  \label{pic:dalek_front}
\end{figure}

The processors integrated into \Dalek are x86 CPUs from both \Intel and \AMD
(\ARM CPUs may be considered in future updates). Some partitions feature
homogeneous multi-core CPUs, while others showcase heterogeneous SoCs that
include performance, efficient, and ultra-low-power cores (referred to by \Intel
as p-cores, e-cores, and LPe-cores, respectively). Most of these heterogeneous
SoCs also include NPU accelerators optimized for efficient inference of deep
neural networks (such as CNNs and LLMs). Typically, such SoCs are not available
in traditional computing centers, making it especially valuable to include them
in a reproducible benchmarking environment. These architectures are generally
optimized for minimal energy consumption while maintaining high performance, and
they may serve as precursors to future computing center designs. A photo of the
cluster is provided in Fig.~\ref{pic:dalek_front}.

Regarding accelerators, \Dalek aims to cover the main architectures from major
vendors by providing \Nvidia, \AMD, and \Intel GPUs. Today, professional-grade
GPUs are extremely expensive, largely due to high demand in the AI market. By
leveraging gaming GPUs, \Dalek makes it possible to prototype algorithms,
including those for deep learning, at a significantly lower cost.

Each partition typically consists of four nodes of the same type. While this is
small compared to the scale of supercomputers, it is sufficient for developing
and testing distributed applications. The network relies on 2.5 GbE interfaces,
which do not match the performance of fiber channel network found in standard
clusters. However, these partitions allow the use of low-cost motherboards and
mini-PCs, keeping the platform affordable and accessible.

In summary, \Dalek is a unique cluster that facilitates the design and the
optimization of future compute intensive applications. The later will need to
take into account many different layers of optimization. Among them, there are
dedicated instructions on a single core, multi-core parallelism with
heterogeneous cores that sometimes even not share the same ISA, and multiple
nodes that can exchange data on a external network. Of course, the same
complexity extends to the memory.

This paper is organized as follow: Sec.~\ref{sec:topo} presents the topology of
the cluster in term of components and network organization, Sec.~\ref{sec:soft}
details the software stack of the front and compute nodes, Sec.~\ref{sec:energy}
introduces a new platform, designed for \Dalek, to precisely measure energy
consumption, Sec.~\ref{sec:benchs} gives synthetic benchmarks run on
heterogeneous components, Sec.~\ref{sec:uses} describes the identified use cases
for both research and education and, finally, Sec.~\ref{sec:conc} concludes this
article.

\section{Cluster Topology}\label{sec:topo}

\begin{figure*}[htp]
  \centering
  \includegraphics[width=1\linewidth]{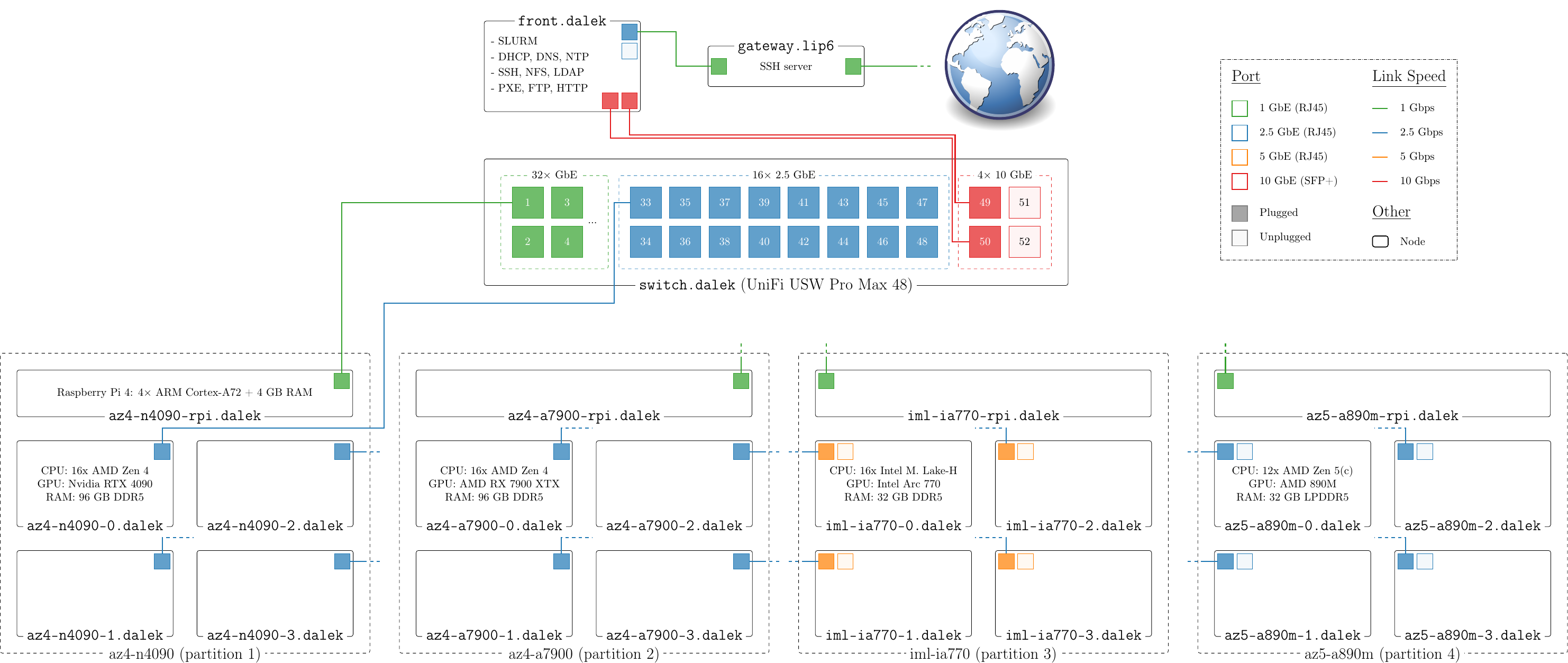}
  \caption{\Dalek topology: four partitions of four compute nodes are connected
    through a 2.5 GbE network and are managed by the frontend node.}
  \label{fig:topology}
\end{figure*}

Fig.~\ref{fig:topology} provides an overview of the cluster topology. \Dalek
consists of four partitions, each containing four nodes, plus a frontend node
(\verb|front.dalek|) responsible, among other tasks, for resource allocation.
A 48-port UniFi USW Pro Max 48 switch (\verb|switch.dalek|) connects all the
compute nodes and the frontend node. The network primarily uses the 2.5 Gigabit
Ethernet protocol (2.5 GbE).

\subsection{Frontend Node}

The front-end node is a \Minisforum mini-PC (MS-01 Work Station) equipped with
four network interfaces: 1) two 2.5 GbE ports, one of which is used to connect
to the gateway, and 2) two SFP+ ports (10 Gbps each) connected to the switch,
providing up to 20 Gbps through link aggregation.

\subsection{Compute Nodes}

The partitions are named according to their CPU and primary GPU. The naming
convention allocates the first three characters to the CPU, followed by a
delimiter (the ``\texttt{-}'' character), and the last five characters to the
GPU. For both the CPU and GPU, the first character indicates the vendor:
``\texttt{a}'' for \AMD, ``\texttt{i}'' for \Intel, and ``\texttt{n}'' for
\Nvidia. The remaining characters specify the processor architecture (for CPUs)
or the product name (for GPUs).

\Dalek's partitions, listed from bottom to top level, are as follows:
\begin{enumerate}
  \item \textbf{az4-n4090} (partition 1) consists of four nodes named
    \texttt{az4-n4090-[0-3]}. Each node is built around a \Minisforum BD790i ITX
    motherboard powered by an \Asus ROG~LOKI~SFX-L~1000W~Platinum Power Supply
    Unit (PSU). Additionally, each node is equipped with an \Nvidia \GeForce
    RTX~4090 GPU connected via the PCI~Express~5.0 bus.
  \item \textbf{az4-a7900} (partition 2) consists of four nodes named
    \texttt{az4-a7900-[0-3]}. Each node uses a \Minisforum BD790i ITX
    motherboard powered by an \Asus ROG~LOKI~SFX-L~1000W~Platinum PSU, with an
    \AMD \Radeon RX~7900~XTX GPU connected to the PCI~Express~5.0 bus.
  \item \textbf{iml-ia770} (partition 3) consists of four nodes named
    \texttt{iml-ia770-[0-3]}. Each node is based on a \Minisforum AtomMan~X7~Ti
    mini-PC. An external \Intel \Arc A770 GPU is connected via the Oculink
    protocol. The external GPU is powered by an \Asus
    ROG~LOKI~SFX-L~1000W~Platinum PSU.
  \item \textbf{az5-a890m} (partition 4) consists of four nodes named
    \texttt{az5-a890m-[0-3]}. Each node is based on a \Minisforum EliteMini
    AI370 mini-PC.
\end{enumerate}

\begin{table}[h!]
\centering
\caption{Specifications of the \Dalek's CPUs, GPUs, SSDs and RAMs.}
\label{tab:specs_cpu_gpu}
{\resizebox{1.0\linewidth}{!}{
\begin{tabular}{l l l l r r r}
\toprule
                  & \multicolumn{6}{c}{\text{Processor (CPU)}}                                                                                                        \\ \cmidrule(lr){2-7}
                  &                      &                                 &                        &                     &                 HW  &                 TDP \\
Partition         & Vendor               & Product Name                    & Architecture           &              Cores  &            Threads  &                 (W) \\
\midrule
\textit{frontend} & Intel                & Core i9-13900H                  &  Raptor Lake-H         &                 14  &                 20  &                115  \\
az4-n4090         & \multirow{2}{*}{AMD} & \multirow{2}{*}{Ryzen 9 7945HX} & \multirow{2}{*}{Zen 4} & \multirow{2}{*}{16} & \multirow{2}{*}{32} & \multirow{2}{*}{75} \\
az4-a7900         &                      &                                 &                        &                     &                     &                     \\
iml-ia770         & Intel                & Core Ultra 9 185H               &  Meteor Lake-H         &                 16  &                 22  &                115  \\
az5-a890m         & AMD                  & Ryzen AI 9 HX 370               &  Zen 5                 &                 12  &                 24  &                 54  \\
\bottomrule
\end{tabular}
}}

\vspace{0.3cm}

{\resizebox{1.0\linewidth}{!}{
\begin{tabular}{l l l l r r r}
\toprule
                  & \multicolumn{6}{c}{\text{Graphical Process Unit (GPU)}}                                                                                           \\ \cmidrule(lr){2-7}
                  &                      &                              &                           &                    &               Shader &                TDP  \\
Partition         & Vendor               & Product Name                 & Architecture              &                SM  &               Cores  &                (W)  \\
\midrule
az4-n4090         & Nvidia               & GeForce RTX 4090             & Ada Lovelace              &               128  &               16384  &                450  \\
az4-a7900         & AMD                  & Radeon 7900 XTX              & RDNA 3                    &                96  &                6144  &                300  \\
iml-ia770         & Intel                & Arc A770                     & Alchemist                 &               512  &                4096  &                225  \\
\addlinespace
\textit{frontend} & Intel                & Iris Xe Graphics             & Raptor Lake GT1           &                96  &                 768  &                 --  \\
az4-n4090         & \multirow{2}{*}{AMD} & \multirow{2}{*}{Radeon 610M} & \multirow{2}{*}{RDNA 2.0} & \multirow{2}{*}{2} & \multirow{2}{*}{128} & \multirow{2}{*}{--} \\
az4-a7900         &                      &                              &                           &                    &                      &                     \\
az5-a890m         & Intel                & Arc Graphics Mobile          & Meteor Lake GT1           &               128  &                1024  &                 --  \\
az5-a890m         & AMD                  & Radeon 890M                  & RDNA 3.5                  &                16  &                1024  &                 --  \\
\bottomrule
\end{tabular}
}}

\vspace{0.3cm}

{\resizebox{1.0\linewidth}{!}{
\begin{tabular}{l l l r l r r r}
\toprule
                  & \multicolumn{3}{c}{\text{Solid State Drive (SSD)}} & \multicolumn{4}{c}{Random Access Memory (\text{RAM})} \\ \cmidrule(lr){2-4} \cmidrule(lr){5-8}
                  &          &                       & Size &        & Size &      & \# of \\
Partition         & Vendor   & Product Name          & (TB) &   Type & (GB) & MT/s &  Chn. \\
\midrule
\textit{frontend} & Samsung  & 990 PRO               &    4 &   DDR5 &  96  & 5200 &     2 \\
az4-n4090         & Samsung  & 990 PRO               &    4 &   DDR5 &  96  & 5200 &     2 \\
az4-a7900         & Samsung  & 990 PRO               &    2 &   DDR5 &  96  & 5200 &     2 \\
iml-ia770         & Kingston & OM8PGP41024Q-A0       &    1 &   DDR5 &  32  & 5600 &     2 \\
az5-a890m         & Crucial  & P3 Plus CT1000P3PSSD8 &    1 & LPDDR5 &  32  & 7500 &     4 \\
\bottomrule
\end{tabular}
}}

\end{table}

\begin{table*}[htp]
\centering
\caption{\Dalek cluster specifications with resource accounting and estimated
  power consumption.}
\label{tab:specs_acc}
{\resizebox{1.0\linewidth}{!}{
\begin{tabular}{c c c c c c r r r r r r r r r r}
\toprule
                                    &                                                               &                           &                                       &                                     &                          & \multicolumn{7}{c}{Amount of Resources}                                                                    & \multicolumn{3}{c}{Power Consumption}       \\ \cmidrule(lr){7-13}\cmidrule(lr){14-16}
Partition                           &                                                               &                           & \multicolumn{3}{c}{GPU}                                                                                &             &          CPU &       CPU HW &           RAM &          iGPU &            dGPU &         VRAM &         Idle &      Suspend &          TDP  \\ \cmidrule(lr){4-6}
Name                                & CPU                                                           & RAM                       & Integrated                            & Discrete                            &                     VRAM &      Nodes  &        Cores &      Threads &          (GB) &         Cores &           Cores &         (GB) &          (W) &          (W) &          (W)  \\
\midrule
\makecell{az4-n4090\\(partition 1)} & \makecell{AMD Ryzen 9 7945HX\\(16x Zen 4 cores)}              & \makecell{96 GB\\DDR5}    & \makecell{AMD Radeon\\610M}           & \makecell{Nvidia GeForce\\RTX 4090} & \makecell{24 GB\\GDDR6X} &          4  &          64  &         128  &          384  &           512 &           65536 &          96  &         212  &           6  &         2100  \\ \addlinespace
\makecell{az4-a7900\\(partition 2)} & \makecell{AMD Ryzen 9 7945HX\\(16x Zen 4 cores)}              & \makecell{96 GB\\DDR5}    & \makecell{AMD Radeon\\610M}           & \makecell{AMD Radeon\\RX 7900 XTX}  & \makecell{24 GB\\GDDR6}  &          4  &          64  &         128  &          384  &           512 &           24576 &          96  &         192  &           6  &         1500  \\ \addlinespace
\makecell{iml-ia770\\(partition 3)} & \makecell{Intel Core Ultra 9 185H\\(16x Meteor Lake-H cores)} & \makecell{32 GB\\DDR5}    & \makecell{Intel Arc\\Graphics Mobile} & \makecell{Intel Arc A770}           & \makecell{16 GB\\GDDR6}  &          4  &          64  &          88  &          128  &          4096 &           16384 &          64  &         260  &          92  &         1360  \\ \addlinespace
\makecell{az5-a890m\\(partition 4)} & \makecell{AMD Ryzen AI 9 HX 370\\(12x Zen 5 cores)}           & \makecell{32 GB\\LPDDR5x} & \makecell{AMD Radeon\\890M}           & --                                  &  --                      &          4  &          48  &          96  &          128  &          4096 &              -- &          --  &          16  &           8  &          216  \\ \addlinespace\midrule
front                               & \makecell{Intel Core i9-13900H\\(14x Raptor Lake-H cores)}    & \makecell{96 GB\\DDR5}    & \makecell{Intel Iris\\Xe Graphics}    & --                                  &  --                      &          1  &          14  &          20  &           96  &           768 &              -- &          --  &          15  &          --  &          115  \\ \addlinespace
*-rpi                               & \makecell{Raspeberry Pi 4\\(4x Cortex-A72 cores)}             & \makecell{ 4 GB\\LPDDR4}  & VideoCore VI                          & --                                  &  --                      &          4  &          16  &          16  &           16  &            -- &              -- &          --  &          12  &          --  &           36  \\ \addlinespace
switch                              & \makecell{Unifi Pro Max 48}                                   & --                        & --                                    & --                                  &  --                      &         --  &          --  &          --  &           --  &            -- &              -- &          --  &          20  &          --  &          100  \\ \addlinespace\midrule
\textbf{Total}                      & --                                                            & --                        & --                                    & --                                  &  --                      & \textbf{21} & \textbf{270} & \textbf{476} & \textbf{1136} & \textbf{9984} & \textbf{106496} & \textbf{256} & \textbf{727} & \textbf{112} & \textbf{5427} \\
\bottomrule
\end{tabular}
}}
\end{table*}

Tab.\ref{tab:specs_cpu_gpu} summarizes the main specifications of the CPUs and
GPUs available in the cluster. Three different CPU models from \Intel and \AMD
are included. The \CoreUltra 9~185H and \Ryzen AI~9~HX370 CPUs feature
heterogeneous core architectures. Specifically, the \CoreUltra 9~185H CPU
contains 2 low-power efficient cores (LPe-cores), 8 efficient cores (e-cores),
and 6 high-performance cores (p-cores), whereas the \Ryzen AI~9HX~370 has 8
e-cores (also known as Zen 5c cores) and 6 p-cores (Zen 5 cores).

The cluster includes six different GPU types: some are discrete with dedicated
memory (VRAM), while others are integrated into the SoC and share unified RAM
with the CPU. This unified memory architecture allows both CPU and GPU to access
the same memory addresses, eliminating the need for costly memory copies.

Tab.~\ref{tab:specs_acc} presents the total number of cores, memory capacity,
and energy consumption of the cluster.

\subsection{Raspberry Pi Nodes}

One Raspberry~Pi~4 with 4~GB of RAM is dedicated to each partition. Each
Raspberry~Pi is responsible for monitoring its corresponding partition in terms
of resource usage and temperature. A visualization system using ARGB LED strips
has been designed to display this information.

\subsection{Network Configuration}


\begin{lstlisting}[style=mystyle, label=lst:network, abovecaptionskip=10pt,
  caption={IPv4 addresses attribution strategy. \texttt{x} stands for the bits
    dedicated to the node addresses.}]
# az4-n4090 partition 1, sub-net: 192.168.1.00/27
# (IP addresses range: [01;030])
11000000 10101000 000000001 000xxxxx
# az4-a7900 partition 2, sub-net: 192.168.1.32/27
# (IP addresses range: [33;062])
11000000 10101000 000000001 001xxxxx
# iml-ia770 partition 3, sub-net: 192.168.1.64/27
# (IP addresses range: [65;094])
11000000 10101000 000000001 010xxxxx
# az5-a890m partition 4, sub-net: 192.168.1.96/27
# (IP addresses range: [97;126])
11000000 10101000 000000001 011xxxxx
\end{lstlisting}

The network organization is divided by partition, as shown in
List.~\ref{lst:network}, where a sub-network strategy is employed. Note that the
subnet masks are virtual; the actual mask is \texttt{255.255.255.0}. Indeed, all
nodes communicate within the same network: \texttt{192.168.1.0/24}.

\begin{table}[b!]
\centering
\caption{Interfaces \& \texttt{192.168.1.0/24} local network.}
\label{tab:network}
{\resizebox{1.0\linewidth}{!}{
\begin{tabular}{l l l r r r}
\toprule
                             &                      &                 &       &              & Switch \\
Host name                    & Interface            & Hardware        &   GbE &        IP @  &   Port \\
\midrule
\texttt{front.dalek}         & \texttt{enp2s0f0np0} & Intel X710      &  10.0 & \texttt{254} &     49 \\
\texttt{front.dalek}         & \texttt{enp2s0f1np1} & Intel X710      &  10.0 & \texttt{254} &     50 \\
\addlinespace
\texttt{switch.dalek}        & --                   & USW Pro Max 48  & 224.0 & \texttt{253} &     -- \\
\addlinespace
\texttt{az4-n4090-0.dalek}   & \texttt{enp5s0}      & Realtek RTL8125 &   2.5 &   \texttt{1} &     33 \\
\texttt{az4-n4090-1.dalek}   & \texttt{enp5s0}      & Realtek RTL8125 &   2.5 &   \texttt{2} &     34 \\
\texttt{az4-n4090-2.dalek}   & \texttt{enp5s0}      & Realtek RTL8125 &   2.5 &   \texttt{3} &     35 \\
\texttt{az4-n4090-3.dalek}   & \texttt{enp5s0}      & Realtek RTL8125 &   2.5 &   \texttt{4} &     36 \\
\texttt{az4-n4090-rpi.dalek} & \texttt{eth0}        & --              &   1.0 &  \texttt{30} &      1 \\
\addlinespace
\texttt{az4-a7900-0.dalek}   & \texttt{enp7s0}      & Realtek RTL8125 &   2.5 &  \texttt{33} &     37 \\
\texttt{az4-a7900-1.dalek}   & \texttt{enp7s0}      & Realtek RTL8125 &   2.5 &  \texttt{34} &     38 \\
\texttt{az4-a7900-2.dalek}   & \texttt{enp7s0}      & Realtek RTL8125 &   2.5 &  \texttt{35} &     39 \\
\texttt{az4-a7900-3.dalek}   & \texttt{enp7s0}      & Realtek RTL8125 &   2.5 &  \texttt{36} &     40 \\
\texttt{az4-a7900-rpi.dalek} & \texttt{eth0}        & --              &   1.0 &  \texttt{62} &      2 \\
\addlinespace
\texttt{iml-ia770-0.dalek}   & \texttt{enp90s0}     & Realtek RTL8157 &   5.0 &  \texttt{65} &     41 \\
\texttt{iml-ia770-1.dalek}   & \texttt{enp90s0}     & Realtek RTL8157 &   5.0 &  \texttt{66} &     42 \\
\texttt{iml-ia770-2.dalek}   & \texttt{enp90s0}     & Realtek RTL8157 &   5.0 &  \texttt{67} &     43 \\
\texttt{iml-ia770-3.dalek}   & \texttt{enp90s0}     & Realtek RTL8157 &   5.0 &  \texttt{68} &     44 \\
\texttt{iml-ia770-rpi.dalek} & \texttt{eth0}        & --              &   1.0 &  \texttt{94} &      3 \\
\addlinespace
\texttt{az5-a890m-0.dalek}   & \texttt{enp99s0}     & Realtek RTL8125 &   2.5 &  \texttt{86} &     45 \\
\texttt{az5-a890m-1.dalek}   & \texttt{enp99s0}     & Realtek RTL8125 &   2.5 &  \texttt{87} &     46 \\
\texttt{az5-a890m-2.dalek}   & \texttt{enp99s0}     & Realtek RTL8125 &   2.5 &  \texttt{88} &     47 \\
\texttt{az5-a890m-3.dalek}   & \texttt{enp99s0}     & Realtek RTL8125 &   2.5 &  \texttt{89} &     48 \\
\texttt{az5-a890m-rpi.dalek} & \texttt{eth0}        & --              &   1.0 & \texttt{126} &      4 \\
\bottomrule
\end{tabular}
}}
\end{table}

Table~\ref{tab:network} summarizes the IP address assignments. In each case,
addresses are assigned contiguously, starting from the first address in the
partition's subnet. The Raspberry Pi is always assigned the last IP address of
the subnet. The frontend node is connected to two switch ports, and these links
are aggregated to achieve up to 20 Gbps throughput.

\section{Software}\label{sec:soft}

\subsection{Operating System}

As with most compute clusters worldwide, Linux has been chosen as the operating
system for \Dalek. Ubuntu 24.04~LTS Server (kernel version 6.08) runs on the
frontend and on most compute nodes, except for the iml-ia770 partition, which
requires a newer kernel (version 6.14) to support 5 GbE and the \Intel \Arc~770
GPU devices.

\subsection{Frontend Services}

\textbf{\texttt{dnsmasq}}.
A combined DHCP and DNS server distributes fixed IP addresses and host names to
the compute nodes based on their MAC hardware addresses. When an unknown
interface contacts the DHCP server, it is assigned an IP address in the range
$[129;159]$. The domain and search domain are set to \texttt{dalek}.

\textbf{\texttt{ufw}}.
Ubuntu includes the Uncomplicated FireWall (UFW), which operates on top of
Netfilter. In \Dalek, UFW is used to perform Network Address Translation (NAT)
for the compute nodes requests. Essentially, all data traffic originating from
the compute nodes destined to Internet is translated: the source address of the
packets is replaced by the frontend's address, and the source port is modified
to encode the original source address. It also serves its main role as a
firewall on all the nodes.

\textbf{\texttt{chrony}}.
An NTP server ensures time synchronization across the entire cluster. The
service itself is synchronized with the LIP6 NTP server (\texttt{ntp.lip6.fr}).
This synchronization is essential for maintaining consistent timestamps
throughout the cluster, which is particularly important for tasks such as
logging and the transactions on the network file system.

\textbf{\texttt{nfs-kernel-server}}.
A Network File System (NFS) is hosted on the frontend node and shared with all
compute nodes. A dedicated 4~TB SSD is used exclusively for the NFS, formatted
with the \texttt{ext4} file system.

\textbf{\texttt{slapd}}.
The Lightweight Directory Access Protocol (LDAP) is used to provide centralized
user account management across the entire cluster. The Domain Component (DC) is
set to \texttt{dalek}, and two Organizational Units (OUs) have been added to the
Directory Information Tree (DIT): \texttt{Users} for individual user accounts
and \texttt{Groups} for user group definitions. Authentication is handled via
the System Security Services Daemon (SSSD), and LDAP is secured using the
Transport Layer Security (TLS) protocol. A self-signed certificate, hosted on
the frontend node, ensures secure logins.

\textbf{\texttt{openssh-server}}.
Finally, the frontend runs a Secure SHell (SSH) server to allow external users
to connect to \Dalek.

\subsection{Compute Nodes Autoinstall}

One of the main challenges is to automate the installation of the compute nodes.
To achieve this, the Ubuntu \texttt{autoinstall} tool was chosen. It relies on a
YAML configuration file provided to the Ubuntu installer. This file can specify
drive partitioning, early and late custom commands, user account creation,
package installation, and more.

To avoid manually plugging a USB key into each node to launch the installer, a
Preboot eXecution Environment (PXE) has been configured using \texttt{dnsmasq}.
All compute nodes are set to boot from the network, and the frontend serves an
Ubuntu image via a lightweight TFTP server (also handled by \texttt{dnsmasq}) to
automatically install the operating system on the local drive. The corresponding
YAML configuration files are hosted on the frontend via an HTTP server
(\texttt{nginx}), and different versions are delivered based on each node's MAC
address, allowing per-partition customization of the OS. This is especially
useful for installing partition-specific drivers, such as those for GPUs.

Finally, thanks to PXE, switching between 1) system installation and 2) booting
from the local drive, can be controlled remotely from the frontend. We measured
that a full (re-)installation of all sixteen compute nodes can be performed
remotely in approximately 20 minutes.

\subsection{SLURM}

Simple Linux Utility for Resource Management (SLURM) is a widely adopted set of
tools for cluster resource management. It is now considered the de facto
standard and is installed on most clusters and supercomputers around the world.

On \Dalek, SLURM version 24.11.3 is deployed, with the \texttt{slurmctld}
service running on the frontend node and the \texttt{slurmd} service running on
each compute node. SLURM is combined with the MUNGE authentication service
(\texttt{munge}), which is used to create and validate credentials. MUNGE is
designed to be highly scalable and secure, making it suitable for HPC
environments.

\textbf{Nodes Powering.}
SLURM provides specific hooks to manage node power states through the
\texttt{noderesume} and \texttt{nodesuspend} scripts. To power on a node, the
system uses the Ethernet Wake-on-LAN (WoL) protocol to send a ``magic packet''
over the network. To power off a node, a dedicated user named
\texttt{powerstate} is automatically created during the compute node
installation. This user is allowed to shut down the node without a password via
specific sudoer rules. Authentication is handled through SSH with public key
only; the corresponding private key is securely held by the \texttt{slurm} user
on the frontend.

The implemented strategy is as follows: nodes are automatically powered off
after 10 minutes of inactivity, and are powered back on when a user submits a
job through SLURM commands such as \texttt{salloc}, \texttt{srun}, or
\texttt{sbatch}. There can be up to a 2-minute delay between the reservation and
the job start, due to the node boot time. As a result, when the cluster is idle,
its energy consumption is extremely low -- estimated at only about 50 watts.

\subsection{Compute Node Settings}

\textbf{Login Policy}.
Users of the compute nodes are provided with a direct SSH access.
SLURM Plug-in Architecture for Node and job (K)control (SPANK) combined with
Linux Pluggable Authentication Modules (PAM) are configured to reject SSH access
to users that have not reserved the resources. In a similar fashion, they are
configured to terminate the shells of the users once their reservation is
expired. This allows for a practical connection to the compute nodes while
preventing interference with legit running jobs.

\textbf{Local Drive}.
The home directories of the users are located on the NFS
(\texttt{/mnt/nfs/users/\{user\_login\}/}). They are thus available on all nodes
but are not suitable for certain tasks such as compilation, due to network
performances issues. To ease such tasks, a semi-permanent local space, called
\emph{scratch}, is available on all compute nodes, on which all users have a
directory. Unlike in traditional compute clusters this local space is not
flushed when a job is terminated. This space is even preserved upon node
re-installations but should not be considered fully permanent. The scratch path
is \texttt{/scratch/\{user\_login\}/} and it is automatically created by SPANK
and PAM at the first login on a compute node.

\textbf{\texttt{proberctl}}. Each compute node runs a specific
\texttt{proberctl} service. The later is in charge of monitoring the node.
For instance, every seconds, \texttt{proberctl} sends the CPU occupancy to its
corresponding Raspberry Pi via SSH. This allows the LED strips to be animated.

\subsection{Unconventional Uses}

We have identified several features that are generally unavailable in
traditional cluster environments. In order to support novel types of
experiments, the following capabilities have been enabled on \Dalek:

\begin{itemize}
  \item Fine-grained control of CPU frequencies using \texttt{cpufrequtils},
  \item Power capping support via \Intel RAPL for CPUs and \texttt{nvidia-smi}
    for \Nvidia GPUs,
  \item Dynamic control of swap file size on SSDs,
  \item Virtual RAM creation on SSDs using \texttt{ndctl} and \texttt{daxctl}.
\end{itemize}

\section{Energy Measurement Platform}\label{sec:energy}

\begin{figure*}[htp]
  \centering
  \subfloat[Main board. Dimensions: 100$\times$75mm.%
    \label{pic:main_board_pcb}]{%
    \includegraphics[width=.5\linewidth]{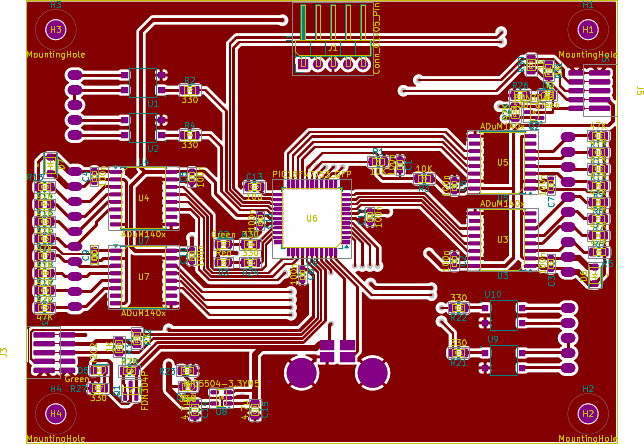}}
  \qquad
  \qquad
  \subfloat[Probe for USB-C. Dimensions: 70$\times$45mm.%
    \label{pic:probe_usbc_pcb}]{%
    \includegraphics[width=.35\linewidth]{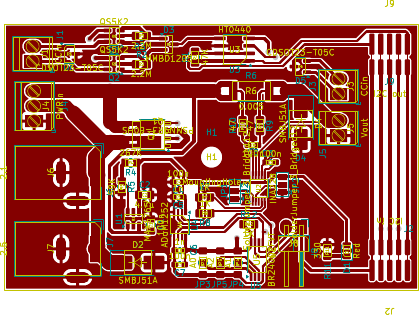}}
  \caption{PCB layout of the energy measurement platform (at the scale).}
  \label{pic:pcb}
\end{figure*}

This section details the hardware platform designed to measure \Dalek's energy
consumption. The goal is to provide a precise, high-frequency power monitoring
system that measures energy usage directly at the power socket. While commercial
solutions do exist, they often lack the flexibility required for research
purposes and are not easily adaptable to custom experimental setups. Our
proposed solution is a modular, open-source platform -- both in hardware and
software -- tailored to meet researchers needs. We also believe this platform
can be reused in other contexts, unlocking new use cases and potentially being
adapted to other clusters in the future. Additionally, since the platform
focuses on socket-level measurements, it complements approaches based on
Model-Specific Registers (MSRs), such as \Intel's RAPL.

The core design is based on separating the platform into two main components:
1) a \textbf{main board}, responsible for aggregating the collected samples and
communicating the measured data to the node, and 2) \textbf{probes}, which
measure voltage and current between the power supply and the compute node. This
architecture allows for designing various probe circuits depending on the power
supply type used by each node, while maintaining standardized interfaces with
the main board. Each compute node is equipped with one main board, and multiple
probes can be connected to it.

\subsection{Main Board}

The PCB layout of the main board circuit is shown in
Fig.~\ref{pic:main_board_pcb}. It is based on a microcontroller from the
\Microchip PIC18 family. The board can aggregate data from up to twelve probes
via two I2C connectors, with up to six probes daisy-chained per connector. The
board is powered through 5~V USB, which also serves as the communication channel
for transmitting the measured samples.

The I2C bus is the primary performance bottleneck, and a maximum sampling rate
of 1000 Samples Per Second (SPS) can be achieved when six probes are connected
to a single bus. Each sample includes the averaged voltage, current, and power
values. Additionally, the number of individual measurements used to compute each
average is reported.

The main board is also equipped with eight GPIOs that can receive binary signals
from the measured compute node. This feature allows for the synchronization of
measurements with specific parts of the running code, which is particularly
useful for fine-grained energy profiling -- such as measuring the consumption of
a specific function or code segment.

\subsection{Probes}

The probes are small and non-intrusive circuits placed between the power supply
and the compute node. They are based on the \Texas INA228 digital power monitor.
While this component supports a maximum sampling rate of 10000 SPS, we have
chosen to reduce it to 4000 SPS in order to enhance measurement resolution down
to the milliwatt level. The reported data consists of averaged values over four
measurements (i.e., 1000 SPS). The main board periodically queries the INA228 to
retrieve and transmit the samples over the I2C bus.

The probe illustrated in Fig.~\ref{pic:probe_usbc_pcb} supports multiple input
types: USB-C and two types of coaxial connectors with different diameters
(2.1$\times$5.5 mm and 2.5$\times$5.5 mm). The USB Power Delivery (PD) 3.1
protocol -- now commonly used in laptops and mini-PCs -- is supported, allowing
power delivery up to 240 Watts.

Another type of probe is planned, specifically designed for PC PSUs. This probe
will connect to the DC outputs of the PSU and will measure power on the 3.3~V,
5~V, and 12~V rails (via Molex, motherboard, CPU, and SATA connectors),
including the new 600~W 12VHPWR connector for GPUs. This design enables more
precise energy measurements for individual components than socket-level
metering, although it excludes the energy consumed by the PSU itself. Multiple
probes will be daisy-chained on the I2C bus to provide per-connector
measurements.

As an additional feature, a dedicated sensor will monitor temperature and
humidity within the cluster environment, although its exact use case remains to
be defined.

\subsection{Application Programming Interface}

It is planned to provide an open source, comprehensive, and well-documented C
API for users. This API will interface with the main board driver to enable the
following functionalities:
\begin{itemize}
  \item Retrieving the measured samples \emph{[available to all users]},
  \item Associating tags to the measured samples via GPIO inputs
    \emph{[available to all users]},
  \item Controlling the power states of the nodes to manually turn power on or
    off \emph{[restricted to administrators]}.
\end{itemize}
To the best of our knowledge, this is the first open hardware and open source
solution combining modular design with high-resolution sampling. For
comparison, the \textsc{Grid'5000} cluster provides around 50~SPS at a
resolution of 0.1~Watt from socket-level measurements
(220 V)\footnote{\textsc{Grid'5000}, ``Raw wattmeters data section'':
\url{https://www.grid5000.fr/w/Energy_consumption_monitoring_tutorial}.}.

\section{Synthetic Benchmarks}\label{sec:benchs}

\subsection{CPU Memory Bandwidth}

\begin{figure*}[htp]
  \centering
  \subfloat[L1d cache, buffer of 16 KB.\label{plot:cpu_throughput_l1d}]{%
   \includegraphics[width=1\linewidth]{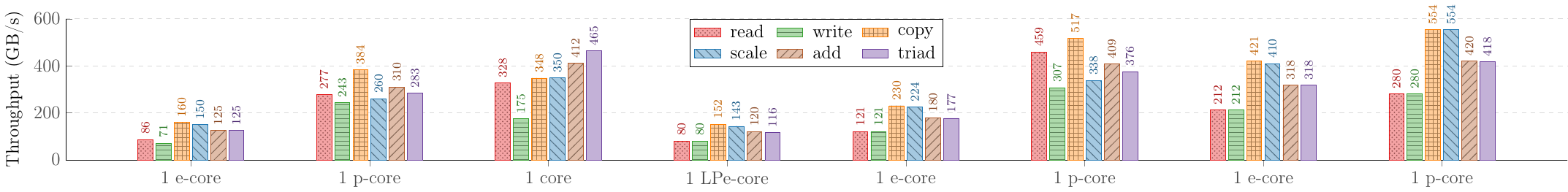}}
  \\
  \vspace{-0.3cm}
  \subfloat[L2 cache, buffer of 512 KB.\label{plot:cpu_throughput_l2}]{%
    \includegraphics[width=1\linewidth]{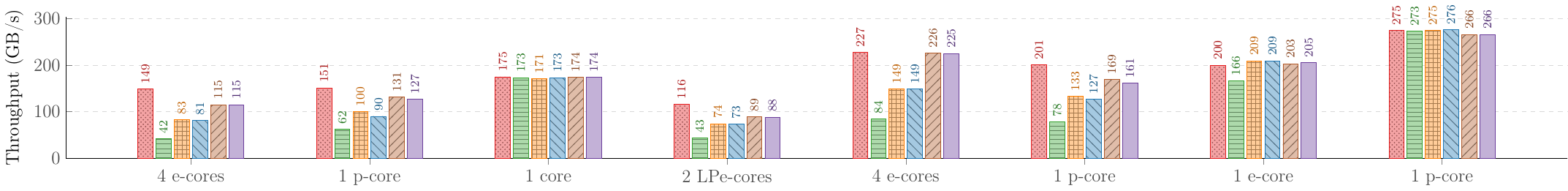}}
  \\
  \vspace{-0.3cm}
  \subfloat[L3 cache, buffer of 20 MB (except for AI~9~HX~370 p-cores where a
    buffer of 10 MB is used).\label{plot:cpu_throughput_l3}]{%
    \includegraphics[width=1\linewidth]{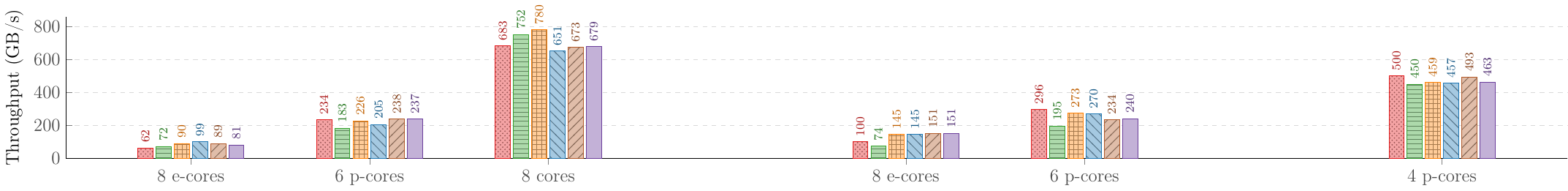}}
  \\
  \vspace{-0.3cm}
  \subfloat[RAM, buffer of 2 GB.\label{plot:cpu_throughput_ram}]{%
    \includegraphics[width=1\linewidth]{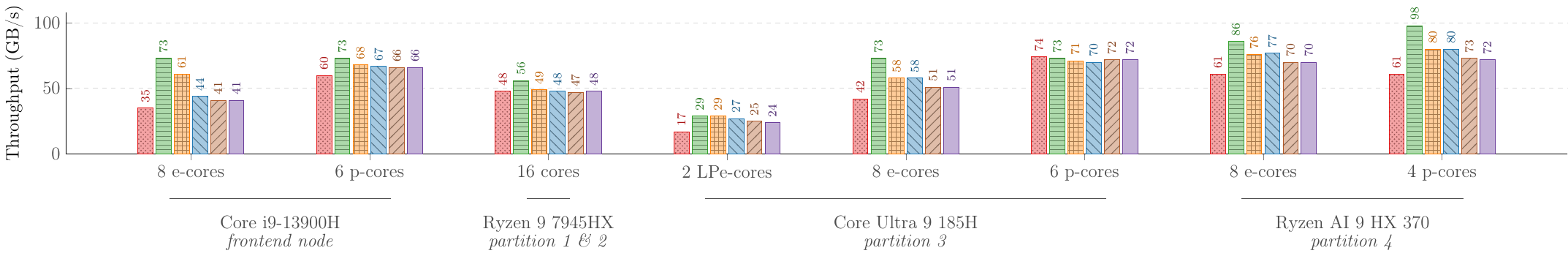}}
  \caption{CPU memory throughput measured with the \texttt{bandwidth}
    benchmark. Higher is better.}
  \label{plot:cpu_throughput}
\end{figure*}

CPU memory throughput is the limiting factor in most real-world applications.
This is why modern CPUs come with multiple levels of cache, from Level 1 (L1,
the smallest but fastest) to Level 3 (L3, the largest but slowest). This section
aims to study the memory throughput of these caches and of the RAM on the \Dalek
processors. For this purpose, the \texttt{bandwidth}
benchmark\footnote{\texttt{bandwidth} repository:
\url{https://github.com/alsoc/bandwidth}.} is used. Developed at the LIP6
laboratory within the ALSOC team, it is inspired by the well-known HPC
\texttt{STREAM} benchmark but offers more flexibility. The benchmark relies on
C++ metaprogramming techniques (templates) and can run on multiple buffer sizes
without requiring recompilation. Additionally, the code supports multithreading
through OpenMP and is explicitly vectorized using intrinsic calls to achieve the
highest possible performance. For example, non-temporal stores are used when
writing data.

The following micro-benchmarks are studied:
\setlist[itemize]{leftmargin=1.5cm}
\begin{itemize}
  \item[\textbf{read}:]   \verb|x = A[i]|
  \item[\textbf{write}:]  \verb|A[i] = x|
  \item[\textbf{copy}:]   \verb|B[i] = A[i]|
  \item[\textbf{scale}:]  \verb|B[i] = x * A[i] |
  \item[\textbf{add}:]    \verb|C[i] = A[i] + B[i]|
  \item[\textbf{triadd}:] \verb|C[i] = x * A[i] + B[i]|
\end{itemize}
\setlist[itemize]{leftmargin=0.675cm}
\texttt{A}, \texttt{B}, and \texttt{C} are buffers allocated on the heap, and
\texttt{i} represents the current index of the outer loop (not shown here). The
buffers are accessed contiguously (i.e., streaming). The buffer size is a
parameter used to target different memory levels: smaller buffers mainly stress
fast caches, while larger buffers generate significant traffic on the RAM bus.

Fig.~\ref{plot:cpu_throughput} shows the achieved throughput depending on
1)~buffer size (subplots (a–d)), 2)~CPU types (x-axis), and 3)~core types (also
on the x-axis). Each time, cores sharing a cache or the RAM are grouped together
to maximize throughput. The L1 cache is always dedicated to a single core, which
is why it is measured on one core only. For the L2 cache, it depends on the
architecture: it can be dedicated to one or more cores. Multiple cores always
share the L3 cache and RAM.

As a general trend, LPe-cores and e-cores are slower than p-cores. For \Intel
CPUs, there is a significant improvement in the L1 cache between the Raptor
Lake-H and Meteor Lake-H architectures. The L2 cache of the latest \AMD Zen~5
architecture outperforms the others. \AMD Zen~4 and Zen~5 CPUs have a much
faster L3 cache compared to \Intel CPUs. On the \CoreUltra~9~185H, LPe-cores do
not have access to the L3 cache, and on the \Ryzen~AI~9~HX~370, the L3 cache
size is equivalent to the combined size of its L2 caches, making its throughput
difficult to measure.

Regarding RAM performance, it is generally balanced around 60--80 GB/s, limited
mainly by DDR5 memory technology. It is worth noting that the \Ryzen~AI~9~HX~370
is paired with fast quad-channel LPDDR5, which explains the slight throughput
improvement.

\subsection{CPU Peak Performance}

\begin{figure*}[htp]
  \centering
  \subfloat[Single-core.\label{plot:cpu_peak_single}]{%
   \includegraphics[width=1\linewidth]{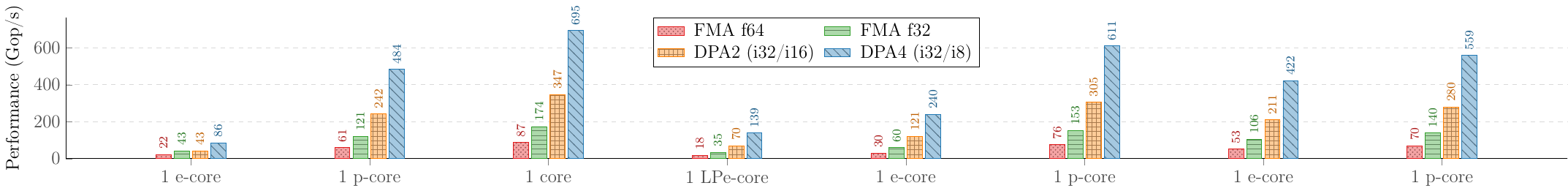}}
  \\
  \vspace{-0.3cm}
  \subfloat[Multi-core grouped by core type.\label{plot:cpu_peak_multi}]{%
    \includegraphics[width=1\linewidth]{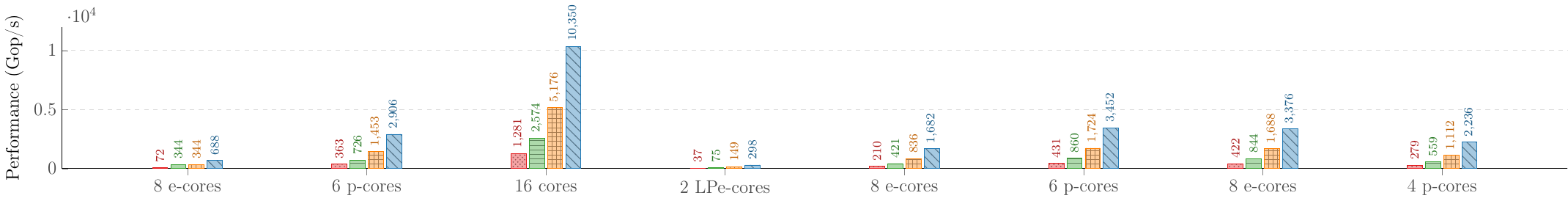}}
  \\
  \vspace{-0.3cm}
  \subfloat[Multi-core accumulated.\label{plot:cpu_peak_multi_acc}]{%
    \includegraphics[width=1\linewidth]{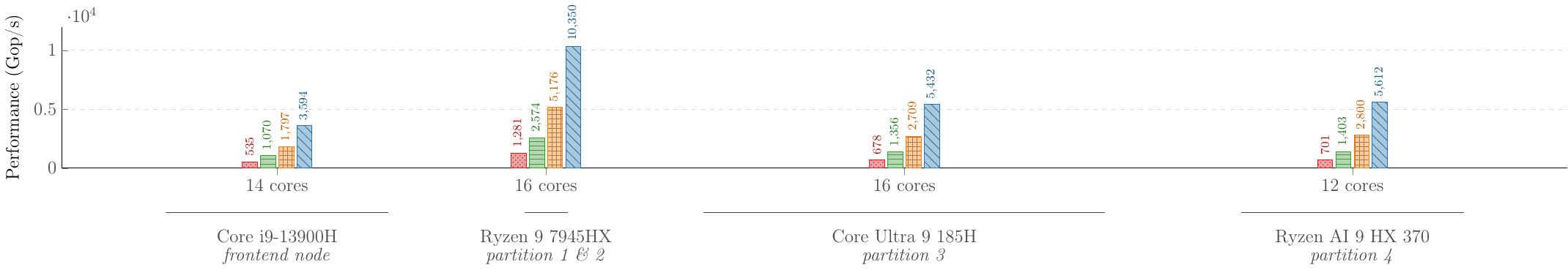}}
  \caption{CPU peak performance in term of number of operations per second.
    Measured with the \texttt{cpufp} benchmark. Higher is better.}
  \label{plot:cpu_peak}
\end{figure*}

In this section, the CPU peak performance in terms of the number of operations
per second (op/s) is measured. Since modern CPUs can execute a large number of
operations, Gop/s (= $10^9$ op/s) are considered. The peak performance is then
evaluated using the \texttt{cpufp} benchmark\footnote{\texttt{cpufp}
repository: \url{https://github.com/pigirons/cpufp}.}. This benchmark consists
of a combination of high-performance instructions without data dependencies,
implemented in assembly code.

On the \Dalek CPUs, the most efficient instructions are:
\setlist[itemize]{leftmargin=1.3cm}
\begin{itemize}
  \item[\textbf{FMA}:] \emph{Fused Multiply-Add} operation that performs
    $d = a \times b + c$ on 32-bit and 64-bit floating-point data format.
  \item[\textbf{DPA2}:] \emph{2-way Dot Product Accumulate} operation that
    performs $c^{i32} = c^{i32} + \sum^2_{s=1}{a^{i16}_s \times b^{i16}_s}$.
    This operation works on 16-bit integers accumulated into a 32-bit integer.
    A variant of this instruction also exists in floating-point where 16-bit
    brain floats (also known as bf16) are multiplied and accumulated into a
    resulting IEEE~754 32-bit float. On the tested architectures, the
    performance of the bf16 version is equivalent to the 16-bit integer version.
  \item[\textbf{DPA4}:] \emph{4-way Dot Product Accumulate} operation that
    performs $c^{i32} = c^{i32} + \sum^4_{s=1}{a^{i8}_s \times b^{i8}_s}$.
    This operation works on 8-bit integers accumulated into a 32-bit integer.
\end{itemize}
FMA instructions have been introduced with the Haswell architecture in 2013,
while DPA2 and DPA4 are more recent (2021 for consumer-grade CPUs). They have
been first introduced in the 512-bit AVX-512 Vector Neural Network Instruction
(AVX-512-VNNI) extension (available starting from the \AMD Zen 4 architecture)
and backported to 256-bit AVX with the AVX-VNNI extension (available starting
from \Intel Alder Lake and \AMD Zen 5 architectures).

Fig.~\ref{plot:cpu_peak} shows the achieved peak performance per CPU type and
core type (x-axis) in single-core, multi-core, and multi-core accumulated modes
((a-c) sub-plots). In single-core mode, the \Ryzen~9~7945HX (Zen~4
architecture) delivers the best performance, even though it is the oldest CPU on
\Dalek. It is also better cooled than the others (large heatsink with a \Noctua
fan) and has a higher TDP, which can explain its strong performance.
Interestingly, the DPA2 instruction does not outperform FMA~f32 on the
\Core~i9-13900H e-core, clearly indicating that this hardware unit is not
implemented on this core type. This changes in the next CPU generation (see the
LPe- and e-cores of the \CoreUltra~9~185H CPU).

In Fig.~\ref{plot:cpu_peak_multi}, the \Ryzen~9~7945HX again outperforms all
competitors, mainly due to its sixteen cores. Indeed, it is the only CPU with
this many performance-class cores. In comparison, the \Core~i9-13900H and the
\CoreUltra~9~185H have six p-cores, while the \Ryzen~AI~9~HX~370 only has four.

Fig.~\ref{plot:cpu_peak_multi_acc} shows the multi-core performance where
LPe-cores, e-cores, and p-cores are combined to estimate the CPU's total peak
performance. The \Ryzen~9~7945~HX achieves about twice the performance of the
\CoreUltra~9~185H and the \Ryzen~AI~9~HX~370, while the \Core~i9-13900H clearly
falls behind.

As a general trend, FMA fp64 performance is doubled by FMA fp32, which is itself
doubled by DPA2, which again is doubled by DPA4 performance. While it might be
tempting to use DPA instructions in algorithms, their specialized nature can
make their integration challenging or even unfeasible in some cases.

\subsection{GPU Memory Bandwidth}

\begin{figure*}[htp]
  \centering
  \includegraphics[width=1\linewidth]{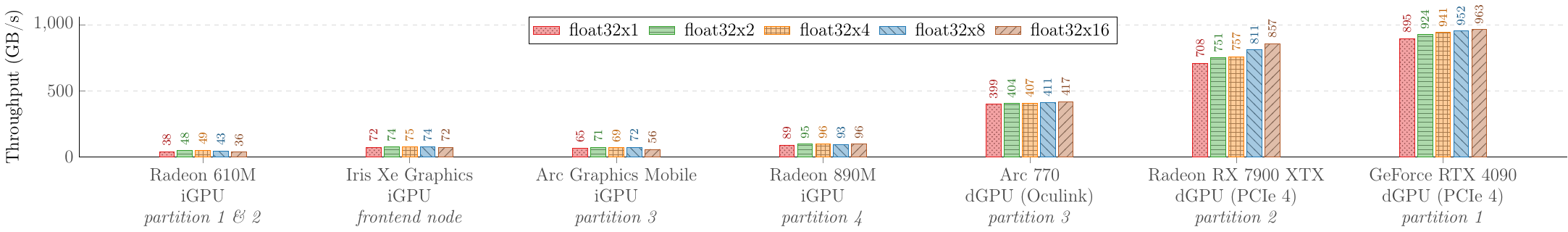}
  \caption{GPU global memory throughput measured with the \texttt{clpeak}
    benchmark (copy kernel). Higher is better.}
  \label{plot:gpu_throughput}
\end{figure*}

The GPU memory bandwidth is measured using the \texttt{clpeak}
benchmark\footnote{\texttt{clpeak}:
\url{https://github.com/krrishnarraj/clpeak}}, which is written in OpenCL. It
performs a copy operation into the global memory. Two different kernels with
distinct access patterns are evaluated to maximize memory throughput, and only
the best result is retained. The global memory refers to VRAM for discrete GPUs
(dGPUs) or system RAM for integrated GPUs (iGPUs).

Various packed data formats are tested, ranging from \texttt{float32x1} (a
single 32-bit float) to \texttt{float32x16} (sixteen 32-bit floats packed
together). Packed data types are known to enhance performance on some GPUs by
helping to hide instruction latency and enabling vector instructions.

Fig.~\ref{plot:cpu_throughput} presents the achieved performance. There is a
clear distinction between iGPUs and dGPUs: in terms of throughput, VRAM is
significantly faster than RAM (up to 10$\times$). For VRAM, packed data types
yield higher performance but within the same order of magnitude, while on iGPUs
using RAM, packed formats seem to have no significant impact.

Cross-comparing results shows that, generally, iGPUs utilize RAM more
efficiently than CPUs. For example, the four p-cores of the \Ryzen~AI~9~HX~370
achieve 80~GB/s (copy kernel), whereas the \Radeon~890M reaches up to 96~GB/s,
representing a notable 20\% improvement over the quad-channel LPDDR5 bandwidth.

\subsection{GPU Peak Performance}

\begin{figure*}[htp]
  \centering
  \includegraphics[width=1\linewidth]{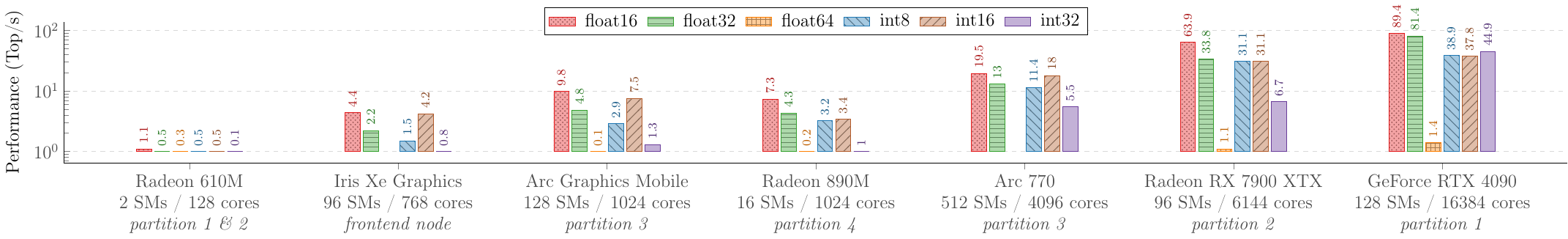}
  \caption{GPU peak performance in term of number of operations per second
    (FMA operations). Only shader cores are considered (tensor and ray tracing
    cores are not evaluated). Measured with the \texttt{clpeak} benchmark.
    Higher is better. Log-scale on the y-axis.}
  \label{plot:gpu_peak}
\end{figure*}

The GPU peak performance is evaluated using the \texttt{clpeak} benchmark. For
floating-point formats, the kernel generates \texttt{mad} instructions without
data dependencies. The \texttt{mad} operation is an approximation of FMA,
prioritizing speed over accuracy. The benchmark also tests integer performance
where classic FMA operations are performed. The evaluated data types include:
\texttt{float16}, \texttt{float32}, \texttt{float64}, \texttt{int8},
\texttt{int16}, and \texttt{int32}.

Fig.~\ref{plot:gpu_peak} shows the GPUs peak performance across different data
types, with the y-axis on a logarithmic scale. The \Radeon~610M, with its two
Streaming Multiprocessors (SMs), is clearly outperformed by others. Other iGPUs
(Iris Xe Graphics, \Arc Graphics Mobile, and \Radeon~890M) reach peak
performances that significantly surpass those of the CPUs. For example, the
\CoreUltra~9~185H CPU reaches up to 5.4 Top/s with the DPA4 instruction, while
the \Arc~Graphics~Mobile GPU delivers 9.8 Top/s using the more flexible FMA on
16-bit floats.

The performance gap between iGPUs and dGPUs is even larger than for memory
bandwidth, nearly an order of magnitude. However, energy consumption should also
be considered when comparing these two GPU types. Typically, iGPUs have a TDP
around 20–30 Watts, whereas \GeForce~RTX~4090 dGPUs can consume up to 450 Watts.

\subsection{GPU Kernel Launch Latency}

\begin{figure}[htp]
  \centering
  \includegraphics[width=1\linewidth]{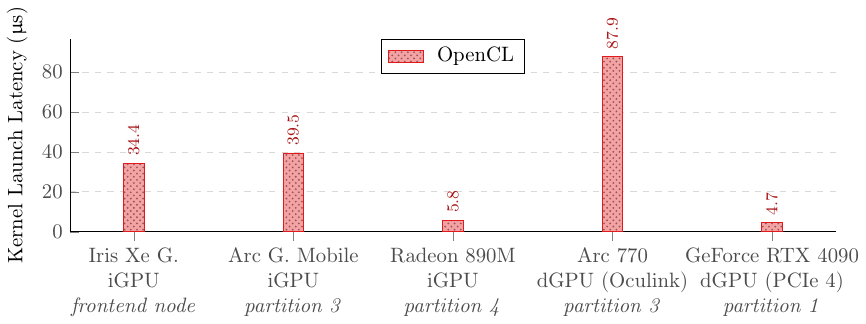}
  \caption{GPU kernel launch latency using the OpenCL API.}
  \label{plot:gpu_kernel_latency}
\end{figure}

Kernel launch latency is the elapsed time between when a GPU kernel is called
from the host (CPU) and when the kernel actually starts executing on the device
(GPU). This latency depends on both software and hardware factors. The software
stack includes the API latency in user space as well as the GPU driver in the
kernel space. On the hardware side, the GPU typically uses a dedicated dispatch
unit to distribute threads across the SMs. This dispatch unit can add extra
latency, especially when the thread grid is very large.

Kernel launch latency is generally not well documented because it is negligible
in typical GPU workloads. However, for applications running small kernels with
frequent communication to the host, this latency can become a limiting factor.
Fig.~\ref{plot:gpu_kernel_latency} shows the measured latency via the OpenCL
API. Note that latency can vary slightly depending on the API used. On the \AMD
\Radeon~610M and \Radeon~RX~7900~RTX, OpenCL event handling is not properly
implemented, which is why their latency values are not shown in the plot.

The \Arc~770 exhibits latency around 90~$\mu$s. It is unclear whether this is
related to Oculink. \Intel iGPUs (Iris~Xe~Graphics and \Arc~Graphics~Mobile)
have latency values around 35--40~$\mu$s, while the \Radeon~890M and the
\GeForce~RTX~4090 have much lower latency, about 5~$\mu$s.


\subsection{SSD Throughput}

\begin{figure}[htp]
  \centering
  \includegraphics[width=1\linewidth]{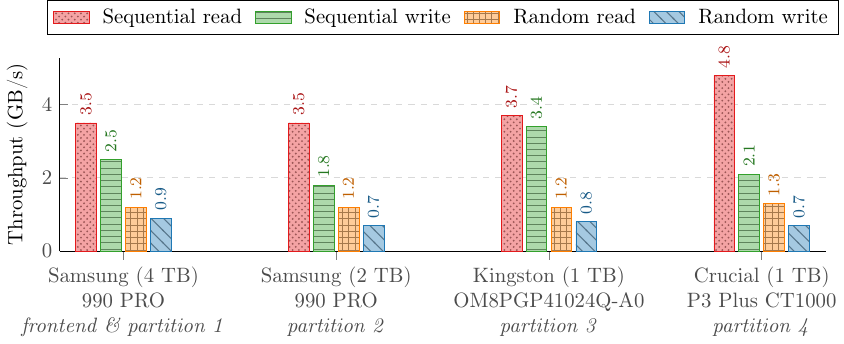}
  \caption{SSD throughput on sequential and random access patterns.}
  \label{plot:ssd_throughput}
\end{figure}

Today, SSDs are game changers compared to the performance of traditional Hard
Disk Drives (HDDs). They enable new use cases; for example, swapping is now
relatively inexpensive compared to the old days. To evaluate the performance of
\Dalek's SSDs, two modes are considered: 1) sequential reads and writes (using
\texttt{dd}) and 2) random reads and writes (using \texttt{iozone}). All tested
SSDs are NVMe drives connected via PCIe 4.0 through M.2 connectors. The drives
use the \texttt{ext4} filesystem. The hardware block size is 512 bytes, while
logical blocks are set to 4096 bytes.

Fig.~\ref{plot:ssd_throughput} shows the achieved throughput depending on the
SSD model and the type of read/write access. As expected, sequential accesses
are about 3 times faster than random accesses, and read accesses are faster than
write accesses. Surprisingly, sequential writes on the \Kingston~OM8PGP4 SSD are
very close in speed to sequential reads.



\section{Use Cases}\label{sec:uses}

The purpose of this section is to identify and report the cluster use cases. It
provides an overview of its impact on scientific research and education.

\subsection{Research}

\textbf{Heterogeneity}.
Some researchers use \Dalek to access specific hardware features. For example,
\emph{D. Orhan and others}~\cite{Orhan2025a} published a paper in HCW'25 on
heterogeneous scheduling across two different types of CPU cores. The validity
of their approach has been partly demonstrated on the iml-ia770 partition, which
includes multiple CPU core types.

This same partition was also used by \emph{M. L{\'e}onardon and
others}~\cite{Leonardon2025} to evaluate the performance of a very fast channel
Polar encoder/decoder for Software-Defined Radio (SDR), within the context of a
challenge organized by ISTC'25.

\textbf{Unconventional Uses}.
\emph{N. Amiot and others} are currently working on Side-Channel Attacks (SCAs),
using formal model-checking to prove the resilience of SoCs and dedicated
cryptographic chips against characterized attackers. \Dalek offers an excellent
trade-off between CPU core frequency (with DFS enabled) and available RAM. This
is especially relevant for the targeted single-threaded symbolic simulations
with memory constraints, where the ability to resize the swap file is helpful.

\textbf{Energy}.
\emph{E. Galvez and others}~\cite{Galvez2025} recently presented preliminary
work at DP2E-AI'25. They used the az5-a890m partition to benchmark CNN
convolution implementations on the \AMD Zen~5 architecture. Both MSRs and socket
power consumption were studied.

\emph{Y. Idouar and others} are currently working on an extension of the work
by \emph{D. Orhan and others}~\cite{Orhan2025a}. One major improvement will be
to incorporate real power consumption from \Dalek platform into the evaluation
of schedulers.

\subsection{Education}

\Dalek offers educational value as it can be used to introduce cluster
environments with a resource manager and batch scheduler. Since \Dalek is small
and resource-constrained, it presents an interesting and challenging environment
for students to learn, similar to what they would encounter on large-scale
clusters.

The ``slow'' network is also noteworthy because it saturates very quickly.
Therefore, even with a small number of nodes, it becomes important to consider
optimizing network communications when designing prototypes. This provides a
great opportunity to introduce MPI compute/communication overlapping.

Many parallel programming paradigms can be explored thanks to \Dalek's
heterogeneity. One example is the SIMD model with large 512-bit vectors using
modern masking techniques. It also allows working with new AI-oriented
instructions (VNNI) and/or the dedicated NPUs included in the latest \Intel and
\AMD SoCs.

Most of \Dalek's CPUs feature a unified memory system, enabling multi-target
zero-copy implementations. Portable APIs like SYCL and Kokkos can take advantage
of this capability.

Vendor-specific APIs such as \Nvidia CUDA, \AMD HIP, and \Intel Level Zero can
also be used to achieve the highest performance on accelerators. These are
especially useful for understanding architectural optimizations.

Finally, there are plans to implement time and energy SLURM quotas (leveraging
the previously introduced energy measurement platform). These additional
constraints will challenge students and provide clear insights into the resource
costs of running simulations. Eco-friendly strategies, such as prototyping on
energy-efficient nodes and cores, will be encouraged.

\section{Conclusion}\label{sec:conc}

This paper presented \Dalek, an experimental and unconventional computing
cluster. Its topology was detailed, along with a series of comprehensive
benchmarks conducted to assess the performance of its heterogeneous components.
A dedicated energy monitoring platform was also designed, enabling a wide range
of experiments on the cluster. Finally, several use cases were presented,
highlighting the relevance and necessity of such a platform for research and
development purposes.

\section*{Online Materials}

\setlist[itemize]{leftmargin=0.5cm}
\begin{itemize}
  \item User documentation: \url{https://dalek.proj.lip6.fr/}
  \item SLURM dashboard: \textit{Coming soon}
  \item Energy Measurement Platform: \textit{Coming soon}
\end{itemize}
\setlist[itemize]{leftmargin=0.675cm}

\section*{Acknowledgment}
{
\noindent
This work has received financial support from the French State, managed by the
French ``Agence Innovation Défense'' (see
\url{https://www.defense.gouv.fr/aid}). Additionally, this work has been
supported by \Minisforum in the form of discounts on mini-PCs and motherboards.
}

\bibliographystyle{IEEEtran}
\bibliography{refs}

\balance






\end{document}